\documentclass[twocolumn,amsmath,amssymb,floatfix,aps,prl,footinbib]{revtex4}

\usepackage{amsmath,amssymb,natbib,bm,graphicx,url}
\usepackage{epstopdf}
\usepackage[british]{babel}

\begin{document}

\title{Pseudo-ballistic transport \\in 3D topological insulator quantum wires}

\author{Joseph Dufouleur$^1$, Louis Veyrat$^1$, Emmanouil Xypakis$^2$, Jens~H. Bardarson$^2$, Christian Nowka$^1$, Silke Hampel$^1$, Barbara Eichler$^1$, Oliver~G. Schmidt$^1$, Bernd B\"uchner$^{1,3}$, and Romain Giraud$^{1,4}$}
\affiliation{
\vspace{5mm}
$^{1}$Leibniz Institute for Solid State and Materials Research, IFW Dresden, D-01069 Dresden, Germany\\
$^{2}$Max-Planck-Institut f\"ur Physik Komplexer Systeme, N\"othnitzer Stra{\ss}e 38, D-01187 Dresden, Germany\\
$^{3}$Department of Physics, TU Dresden, D-01062 Dresden, Germany\\
$^{4}$CNRS - Laboratoire de Photonique et de Nanostructures, Route de Nozay, F-91460 Marcoussis, France\\
}

\date{\today}


\begin{abstract}
Quantum conductance fluctuations are investigated in disordered 3D topological insulator quantum wires. Both experiments and theory reveal a new transport regime in a mesoscopic conductor, pseudo-ballistic transport, for which ballistic properties persist beyond the transport mean free path, characteristic of diffusive transport. It results in non-universal conductance fluctuations due to quasi-1D surface modes, as observed in long and narrow Bi$_2$Se$_3$ nanoribbons. Spin helical Dirac fermions in quantum wires retain pseudo-ballistic properties over an unusually broad energy range, due to strong quantum confinement and weak momentum scattering. 

\vspace{5mm}
\end{abstract}


\maketitle

Disorder strongly alters the nature of charge transport in nanostructures, inducing a crossover from ballistic to diffusive transport when impurities are introduced into a mesoscopic conductor \cite{Akkermans2007}. The transport regime can be identified by quantum coherent transport measurements. When the phase coherence length $L_\varphi (T)$ of quasi-particles becomes comparable to or larger than the length $L$ of a conductor, quantum interference modifies the conductance, and the statistical properties of conductance fluctuations (CF) depend on the nature and the dimensionality of charge transport \cite{Akkermans2007}. For a disordered conductor in the metallic diffusive regime \cite{Stone1985,Altshuler1985,Lee1987}, the variance of conductance fluctuations is a universal constant for $L_\varphi (T) > L$ (fully coherent transport). For a ballistic conductor conductance fluctuations are not necessarily universal \cite{Asano1996}, but they become universal when the dynamics is chaotic \cite{Jalabert1990,Avishai1990a}. If a small amount of impurities is introduced into a ballistic conductor, momentum scattering is generally efficient enough to generate chaos, even in otherwise integrable systems. Therefore, disorder rapidly drives a mesoscopic conductor into the diffusive regime, for which even a change in the location of a single impurity induces universal CF \cite{Cahay1988}. In most cases, momentum scattering is indeed strong enough that diffusive transport already occurs for a conductor length $L \gtrsim l_\text{e}$, where $l_\text{e}$ is the elastic mean free path. In rare cases, momentum scattering is weak and this transition happens for $L \gtrsim l_\text{tr} \gg l_\text{e}$, where the transport mean free path $l_\text{tr}$, typical of backscattered trajectories, becomes the relevant length scale for diffusive transport \cite{Akkermans2007}.\\

\vspace{15mm}

The study of quantum interference in nanostructures of strong 3D topological insulators (3D TI) already revealed the specific nature of quasi-particles at their surface. A clear evidence of metallic surface states was given by the demonstration of periodic Aharonov-Bohm oscillations in Bi$_2$Se$_3$ nanowires \cite{Peng2010}, and Berry phase effects were predicted in 3D TI quantum wires \cite{Ostrovsky2010,Rosenberg2010,Bardarson2010,Zhang2010a,Bardarson2013} and reported in Bi$_2$Se$_3$ nanostructures \cite{Hong2014}. The unusual scattering behavior of helical Dirac fermions, initially found by scanning tunnelling microscopy \cite{Roushan2009,Zhang2009c,Seo2010} and theory \cite{Culcer2010}, was confirmed by a study of decoherence in disordered Bi$_2$Se$_3$ quantum wires \cite{Dufouleur2013}. 
It results from the spin-momentum locking of Dirac fermions at the surface of a 3D TI, which favors enhanced forward scattering and therefore gives a much longer transport length than the elastic mean free path $l_\text{e}$ \cite{Culcer2010}. 
In a narrow nanostructure with perimeter $L_\text{p} \lesssim l_\text{tr}$, the \emph{transverse} motion of helical Dirac fermions is quantized, and its ballistic nature can be directly revealed by the temperature dependence of the amplitude of Aharonov-Bohm oscillations \cite{Dufouleur2013}. 
In this work, we report on a new mesoscopic transport regime evidenced in the \emph{longitudinal} motion of helical Dirac fermions in disordered 3D topological insulator quantum wires, for which ballistic properties persist beyond the transport mean free path, characteristic of diffusive transport. This \emph{pseudo-ballistic} regime is evidenced by non-universal conductance fluctuations in long Bi$_2$Se$_3$ nanowires in the limit $L \ge l_\text{tr}$. 
Theory further reveals that spin helical Dirac fermions in quantum wires retain pseudo-ballistic properties over an unusually broad energy range, due to strong quantum confinement and weak momentum scattering. 


\begin{figure*}[!t]
\includegraphics[width=2.0\columnwidth]{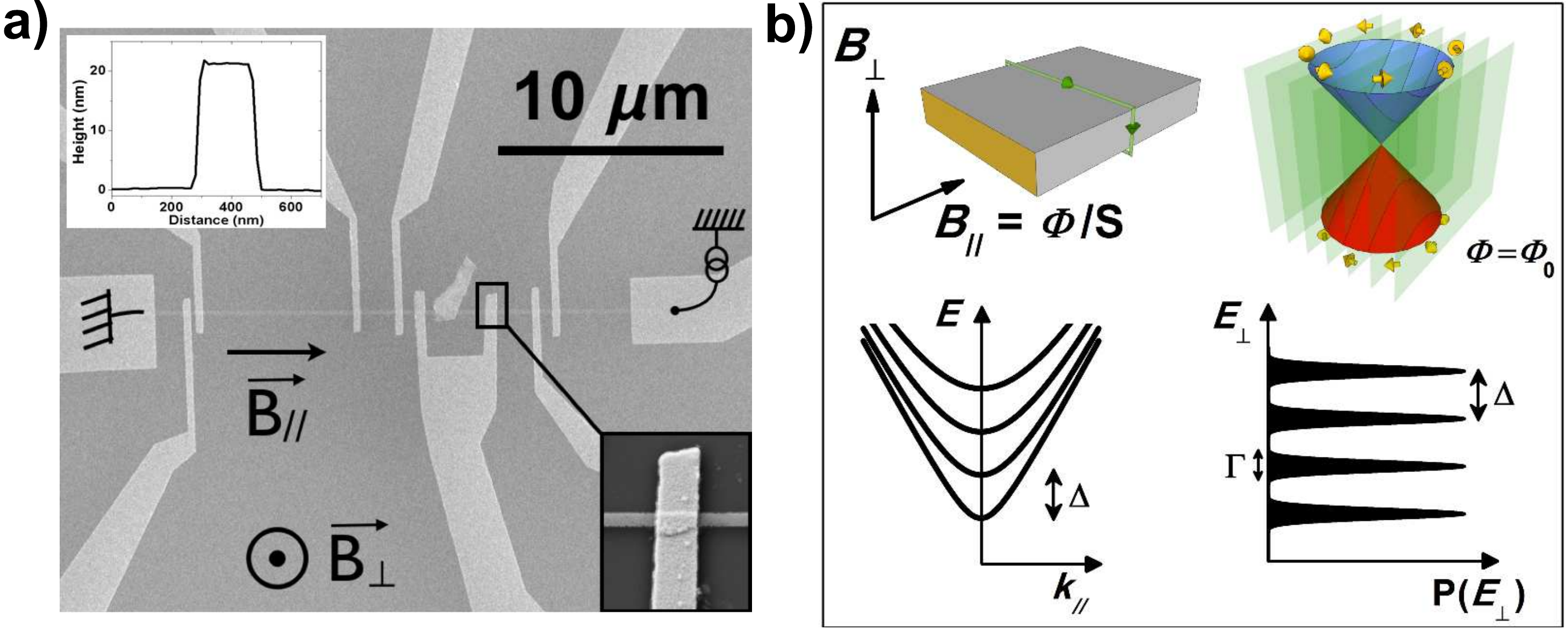}
\caption{
\textbf{a}, Scanning electron microscope image of a Bi$_2$Se$_3$ nanoribbon (width $w=170$~nm, height $h=20$~nm and perimeter $L_\text{p}=380$~nm), contacted with CrAu leads. Three mesoscopic conductors are measured, with different lengths $L_1\approx400$~nm, $L_2\approx1\mu$m or $L_3\approx6\mu$m. Inset: Profile of the atomically flat structure measured by atomic force microscopy. 
\textbf{b}, Schematics of the measurement geometry in magnetic field and of the 1D energy modes in presence of quantum confinement. The transverse energy spacing $\Delta$ remains larger than the disorder broadening $\Gamma$, even in the diffusive limit ($L \ge l_\text{tr}$). 
}
\label{fig1}
\end{figure*}

\newpage

Conductance fluctuations of a long and narrow mesoscopic Bi$_2$Se$_3$ nanoribbon grown by the vapor transport method were investigated at very low temperatures in a $^3$He/$^4$He dilution refrigerator and studied for different lengths between ohmic CrAu contacts (Fig.~\ref{fig1}a). Due to a high density of Se vacancies in the bulk, such a quantum wire is metallic, with the surface Fermi energy $E_\text{F}$ being about 250~meV above the Dirac point. The total conductance results from a comparable contribution from surface and bulk carriers. 
However, conductance fluctuations are dominated by quasi-1D quantized surface modes, as shown later. In long wires, the relative contribution of bulk states to CF is indeed reduced by at least an order of magnitude, due to their much shorter phase coherence length. For helical Dirac fermions, using $v_\text{F}\approx 5\cdot 10^5$~ms$^{-1}$ as obtained from angle-resolved photoemission spectroscopy \cite{Hsieh2009,Kuroda2010,Kordyuk2012}, we find $l_\text{tr} \approx 300$~nm and about $N \approx 2E_\text{F}/\Delta \approx 80$ transverse modes, with an energy level spacing $\Delta = hv_\text{F}/L_\text{p} \approx 6$~meV for a perimeter $L_\text{p}=380$~nm. As sketched in Fig.~\ref{fig1}b), these electronic modes are dispersive in the longitudinal direction and their spin texture in zero field is reminiscent of that of the helical Dirac cone in absence of quantum confinement. 
Importantly, their transverse energy is a periodic function of the longitudinal field $B_{\parallel}$ through an Aharonov-Bohm phase. In contrast, a transverse magnetic field $B_{\perp}$ has little influence on the high-energy spectrum of Dirac fermions (see Supplementary Information) and can therefore be conveniently used to probe aperiodic conductance fluctuations. 
Due to the weak momentum scattering of helical Dirac fermions by impurities, an important and unusual property of the quantized energy spectrum of 3D TI quantum wires is that the transverse energy level spacing $\Delta$ remains larger than the disorder-induced broadening $\Gamma$, even for a conductor length which exceeds the transport length.

\begin{figure*}[!t]
\includegraphics[width=2.0\columnwidth]{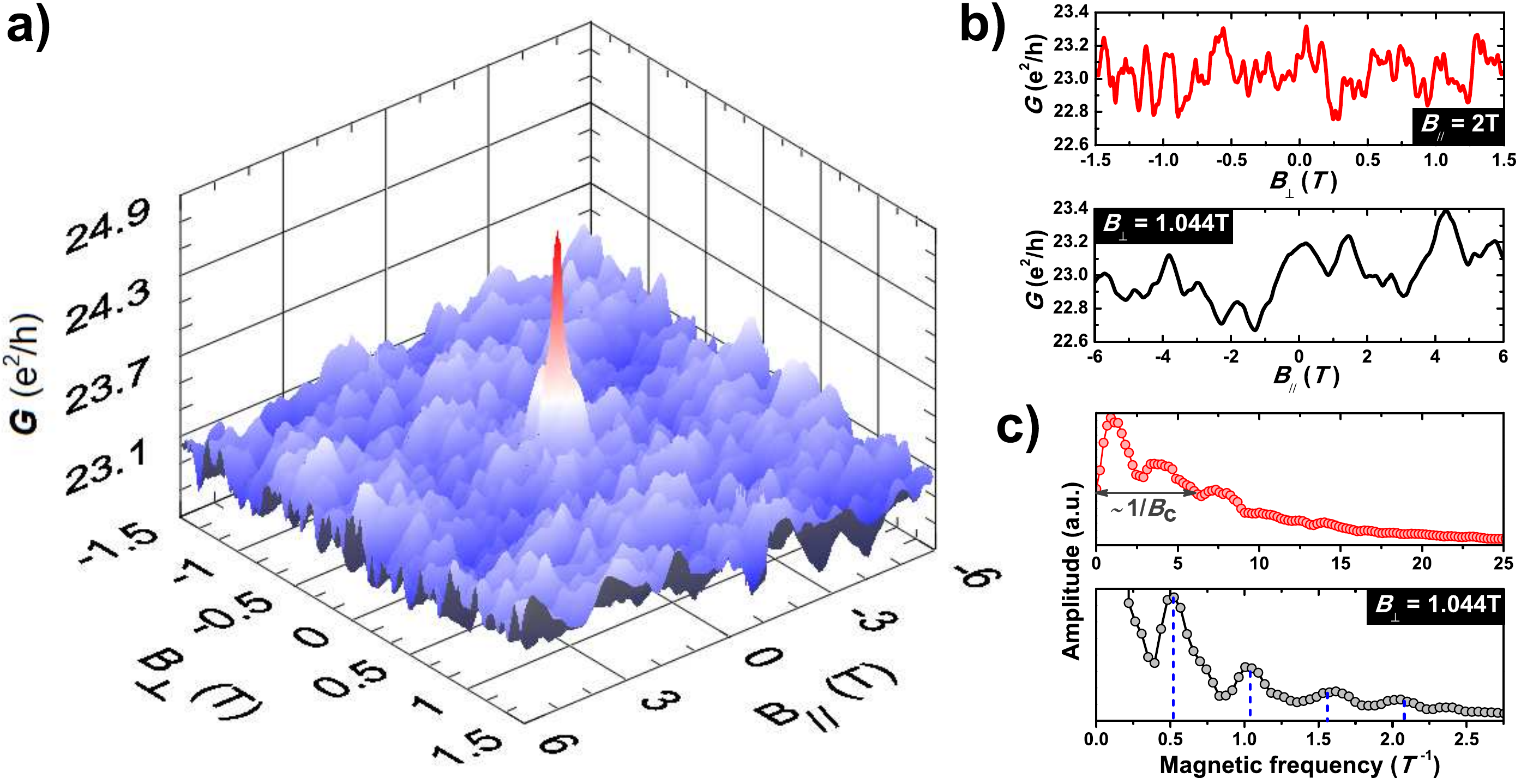}
\caption{
\textbf{a}, Longitudinal and transverse magneto-conductance $G(B_{\parallel},B_{\perp})$ of the quantum wire with length $L_2=1~\mu$m, for which $l_\text{tr}\approx L_\text{p}<L_2<L_{\varphi}$, measured at $T=30$~mK. 
\textbf{b}, Details from a): \emph{top}, $G(B_{\perp})$ for a constant $B_{\parallel}=2$~T (aperiodic conductance fluctuations), and \emph{bottom} $G(B_{\parallel})$ for a constant $B_{\perp}=1.044$~T (periodic Aharonov-Bohm oscillations). 
\textbf{c}, Fast-Fourier transforms of magneto-conductance traces shown in b). Dashed lines refer to four AB harmonics. The horizontal arrow shows the half-width at half-maximum for conductance fluctuations, which relates to the 200~mT correlation field seen in b),\emph{top}. 
}
\label{fig2}
\end{figure*}

\begin{figure*}[!t]
\includegraphics[width=2.0\columnwidth]{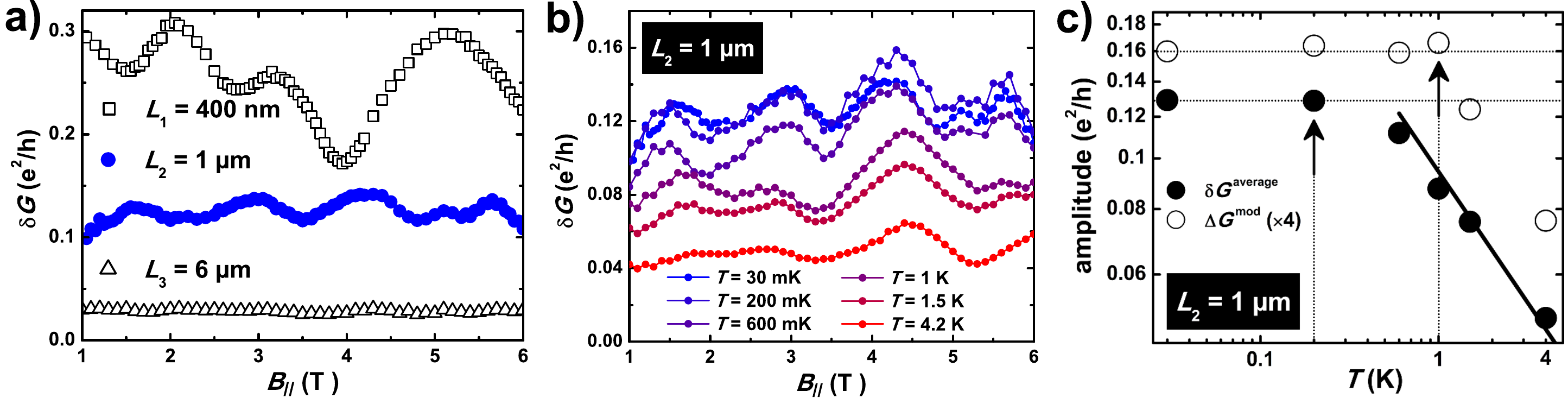}
\caption{
\textbf{a}, Longitudinal-field dependence of the standard deviation of conductance fluctuations $\delta G$ measured at very low temperatures for three different lengths, with $l_\text{tr}\approx L_1 <L_{\varphi}$, $l_\text{tr}<L_2\approx L_{\varphi}$ and $l_\text{tr}<L_{\varphi}<L_3$. Non-universal conductance fluctuations show a $\Phi_0$-periodic modulation, corresponding to a 1.6T period in $B_{\parallel}$.
\textbf{b}, Longitudinal-field dependence of $\delta G$ for the intermediate length $L_2=1~\mu$m, with $L_2/l_\text{tr}\approx3$, measured at different temperatures.
\textbf{c}, Temperature dependence of the amplitude of conductance fluctuations and of their maximum peak-to-peak modulation by a longitudinal field, as inferred from b). Dotted lines are guides for the eye. Thick line: $1/\sqrt{T}$ dependence of $\delta G(T)$ in the regime $L_{\varphi}(T)<L_2$.
}
\label{fig3}
\end{figure*}

Using a 3D-vector superconducting magnet, we measured the full mapping of quantum corrections to the classical conductance of a mesoscopic Bi$_2$Se$_3$ nanoribbon with different lengths $L$ in the limit $L \ge l_\text{tr}$, as shown in Fig.~\ref{fig2}a) for a wire length $L=1~\mu$m at $T=30$~mK. Quantum interference of topological surface states results in a sharp weak anti-localization peak around zero field, and in different field dependences (Fig.~\ref{fig2}b) depending on whether the magnetic field is swept along the wire axis (periodic Aharonov-Bohm oscillations) or perpendicular to it (aperiodic conductance fluctuations). The distinct nature of the quantum interference patterns in finite fields is clearly evidenced by the Fourier transform of magneto-conductance traces measured in a longitudinal or perpendicular field (Fig.~\ref{fig2}c), revealing their periodic and aperiodic characteristics, respectively, and their very different correlation fields. 
Note that the amplitude of CF due to bulk states always remains negligible in the long-wire limit $L \gg L_\varphi^{BS}$ considered here. 
Indeed, the dominant contribution of helical Dirac fermions to conductance fluctuations is unambiguously confirmed by their longitudinal-field dependence (Fig.~\ref{fig3}a). Surprisingly, and contrary to expectations in the fully-coherent regime for diffusive transport ($L_{\varphi} \ge L \ge l_\text{tr}$), the standard deviation of conductance fluctuations saturates at low temperatures to a value $\delta G^{\text{sat}}$ that varies with the longitudinal magnetic field. A striking feature is the \emph{periodic modulation} of $\delta G^{\text{sat}} (B_{\parallel})$, a hallmark of surface-state transport. All segments show a 1.6~T period in $B_{\parallel}$, which corresponds to a $\Phi_0$ periodicity in flux, in agreement with the electrical cross-section of surface states taking surface oxidation into account \cite{Dufouleur2013}. 
The amplitude of the modulation is reduced with increasing the length $L$, whereas its damping with temperature is unchanged. Remarkably, non-universal conductance fluctuations remain visible even in the longest conductor studied with $L=6~\mu$m, for which $L/l_\text{tr}\approx$~20 (see Supplementary Information).

 
The analysis of the temperature dependence of conductance fluctuations reveals two further important properties, which are the diffusive nature of the longitudinal motion and the disorder-induced broadening of quantized transverse modes. 
Since the phase coherence length of helical Dirac fermions is much longer than the transverse dimension of the nanowire, quantum coherent transport occurs in the quasi-1D limit. At high enough temperature, $L_{\varphi}(T)<L$ and the size averaging of CF varies as\cite{Lee1987,Akkermans2007}~~$\delta G(T)\approx\delta G^{\text{sat}} [L_{\varphi}(T)/L]^{\frac{3}{2}}$. In the diffusive regime and for decoherence induced by electron-electron interactions \cite{Altshuler1982,Akkermans2007}~~$L_{\varphi}(T) \propto 1/T^{\frac{1}{3}}$, which gives $\delta G(T) \propto 1/\sqrt{T}$ in agreement with our measurements (Fig.~\ref{fig3}c). This suggests a change in the nature of decoherence for the longitudinal and transverse motions in our quantum wires, due to different charge transport regimes (pseudo-ballistic \emph{vs}. ballistic). 
At very low temperature, the standard deviation of conductance flcutuations saturates at $\delta G^{\text{sat}}$ when $L_{\varphi}(T)\geq L$, a crossover that occurs at $T\approx200$~mK for the 1~$\mu$m long conductor. 
Strikingly, the modulation of $\delta G^{\text{sat}}$  in a longitudinal field is temperature independent up to about $T^{\ast}=$1~K. This is seen in Fig.~\ref{fig3}b,c) for the intermediate length $L=1~\mu$m ($L/l_\text{tr}\approx$~3), but the same behavior was found for all wire lengths. It shows that the amplitude of non-universal CF is not directly related to $L_{\varphi}$, which was already suggested by the length dependence. The crossover observed at $T^{\ast}$ rather corresponds to the limit when the thermal broadening of quantized levels compares to their disorder broadening. Therefore, this measure gives a direct access to the strength of disorder broadening, and we infer $\Gamma \approx 4\times$k$_\text{B}T \approx 0.4$~meV. This value remains much smaller than the energy level spacing $\Delta \approx 6$~meV, a necessary condition for pseudo-ballistic transport to occur. 

The existence of non-universal conductance fluctuations in a diffusive conductor is at odds with standard theories for massive quasi-particles \cite{Stone1985,Altshuler1985,Lee1987} and for Dirac fermions \cite{Kechedzhi2008,Kharitonov2008}, which predict universal values of $\delta G$ in the fully-coherent diffusive regime. Although reduced or enhanced conductance fluctuations can occur for a ballistic conductor with a small number of quantized modes \cite{Tamura1991,Higurashi1992,Nikolic1994,Asano1996}, universality is restored even by a small disorder if the number of transverse modes exceeds a few units \cite{Grincwajg1996}. Similar results were found for Dirac fermions in presence of a strong long-range disorder and without quantum confinement \cite{Rossi2012}. 
In the long Bi$_2$Se$_3$ quantum wires studied here, both conditions $L \ge l_\text{tr}$ and $G\gg G_0$ (number of populated transverse modes $N\gg1$) should therefore set the system in the universal regime. This is clearly not the case. In a strong 3D TI quantum wire, this situation actually corresponds to a pseudo-ballistic transport regime that originates from the coexistence of the quasi-1D ballistic nature of charge transport (quantized transverse momentum) with the ergodic evolution of the longitudinal momentum, typical of the diffusive limit.

\begin{figure*}[!t]
\includegraphics[width=2.0\columnwidth]{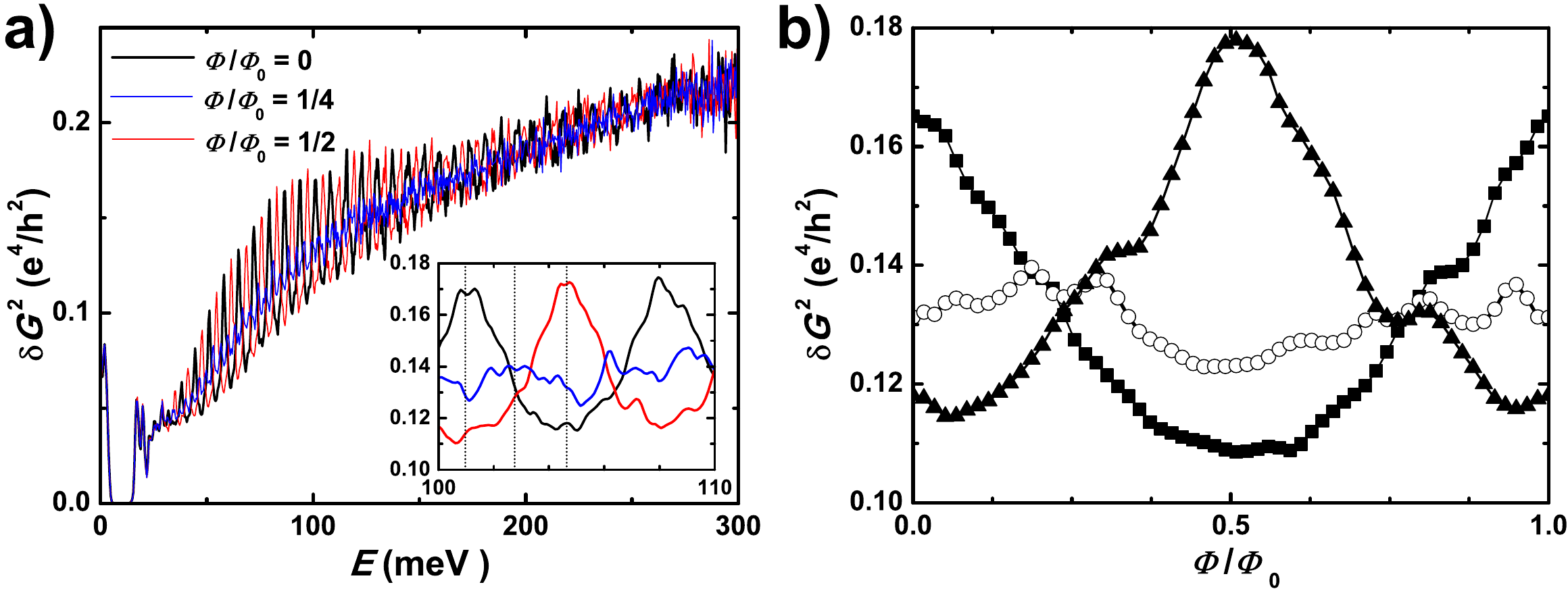}
\caption{
\textbf{a}, Energy dependence of the variance of conductance fluctuations in a disordered 3D topological insulator quantum wire with dimensions similar to experiments (width $w=120$~nm, height $h=20$~nm and length $L=350$~nm) and a correlation length of disorder of 10~nm, typical of disordered Bi$_2$Se$_3$. Calculations are done for a constant transverse field $B_{\perp}=1$~T and three different longitudinal fields which correspond to a magnetic flux $\Phi=n\Phi_0$ ($n=0,\frac{1}{4},\frac{1}{2}$). The inset shows a zoom in a reduced energy window.   
\textbf{b}, Flux dependence of the variance for three different energies shown as dotted lines in a): $E_0=101$~meV ($\blacksquare$), $E=E_0+\frac{1}{4}\Delta$ ($\bigcirc$) and $E=E_0+\frac{1}{2}\Delta$ ($\blacktriangle$), with $E_0 \approx E_{(n=14;\Phi/\Phi_0=\frac{1}{2})}$ and $\Delta=7.2$~meV. When a conductance channel opens, non-universal conductance fluctuations are modulated by the magnetic flux. 
}
\label{fig4}
\end{figure*}

Based on scattering matrix formalism, we show that this unique and new behavior in mesoscopic transport results from a combination of both the even energy spectrum of confined Dirac fermions and their enhanced forward scattering by disorder, due to their spin chirality. To theoretically model our experiments we adapt a continuous Dirac fermion description of the surface state \cite{Bardarson2007a,Bardarson2010} and take the bulk to be an inert insulator (see Supplementary Information for details). For a fixed chemical potential $\mu$, the number of propagating modes in the nanowire is $N = 2\mu/\Delta$. Since in our experiments $N$ is generally large due to the pinning of the Fermi energy in the conduction band of Bi$_2$Se$_3$, we studied the evolution of conductance fluctuations over a broad energy range, and our results reveal that their ballistic nature persist in the large-$N$ limit. The statistics of conductance fluctuations are obtained from sampling over many different microscopic configurations of disorder ($\sim 1000$) and calculated for three different values of the flux $\Phi/\Phi_0=(0,\frac{1}{4},\frac{1}{2})$, which correspond to different configurations of the quantized energy spectrum. As shown in Fig.~\ref{fig4}a) for a fixed perpendicular field $B_\perp = 1$~T, we reveal the energy dependence of non-universal conductance fluctuations. Close to the charge neutral Dirac point ($E = 0$), the physics is dominated by a chiral mode arising from quantum Hall physics \cite{Lee2009,Juan2014,Ilan2014a}. As the energy is increased, quantized transverse modes are weakly perturbed by the transverse field and a clear oscillating behavior of the variance of the conductance is observed, which corresponds to the opening of discrete transport channels. Since the exact value of the chemical potential at which new channels open depends on the parallel field, these oscillations are shifted by the flux. As a consequence, non-universal conductance fluctuations can be probed at a fixed energy by measuring the flux dependence of the variance.  This is shown in Fig.~\ref{fig4}b) for three different energies, which correspond to either the opening of a conductance channel ($E=101$~meV and $E=104.6$~meV) or to nearly closed or opened channels ($E=102.8$~meV). The modulation of the variance is indeed periodic in flux, with a period that corresponds to one flux quantum through the cross section of the quantum wire, and harmonics are clearly present, as expected in the fully coherent regime. Importantly, the amplitude of the modulation can be nearly as large as the variance itself. At larger energies, the modulation is reduced due to the very large number of modes, but it remains significant even at $E=250$~meV, for which both the variance of the conductance and the amplitude of its modulation compare reasonably well with our experiments.

In presence of quantum confinement, a key difference between helical Dirac fermions and other quasi-particles lies in the relative size of the level spacing $\Delta$ and the disorder broadening $\Gamma$ of transverse modes in a nanowire. In general, $\Gamma \gg \Delta$ so that all energy levels overlap. The discretization is washed out and conductance fluctuations are universal. In this limit, there would be no difference between a topological insulator and a charge-accumulation layer at the surface of a semiconductor with strong spin-orbit coupling. This regime corresponds to the case of both metallic and semiconducting nanowires, for which $\Delta$ is small and a small disorder induces a rather large broadening $\Gamma$ of quantized energy levels. In the opposite limit considered here, $\Gamma < \Delta$, a new pseudo-ballistic transport regime emerges, for which signatures of the discreteness and the Dirac nature of the spectrum can be probed even in the presence of a large number of modes. In 3D topological insulator nanowires, this transport regime is surprisingly robust and spans over a significantly large energy range. This is due to the weak scattering of spin helical Dirac fermions by disorder, which limits disorder broadening, and to both the increased energy quantization of Dirac fermions with respect to massive quasi-particles and the regular distribution of quantized energies in a cylinder geometry.

In conclusion, non-universal conductance fluctuations were observed in the diffusive regime of highly-doped Bi$_2$Se$_3$ nanoribbons. Their variance shows a periodic modulation with magnetic flux, which is a hallmark of surface-state transport. We reveal that quasi-1D modes retain their ballistic nature over distances which exceed the transport length typical of diffusive transport, due to weak scattering by disorder, and over a broad energy range, due to both periodic boundary conditions in a hollow geometry (regular distribution of quantized energy levels) and to the large level separation for Dirac fermions in presence of quantum confinement. Unlike massive quasi-particles, Dirac fermions can thus give rise to exotic situations in which ballistic and diffusive properties become entangled. Previously, signatures of diffusion were found in ballistic graphene nanostructures at the Dirac point \cite{Tworzydlo2006,Miao2007,DiCarlo2008,Danneau2008}, corresponding to \emph{pseudo-diffusive} transport \cite{Asboth2009}. In analogy, signatures of ballistic transport in a diffusive conductor correspond to \emph{pseudo-ballistic} transport. This novel transport regime in a mesoscopic conductor could open new directions of research in mesoscopic physics unexplored to date. Besides, the ballistic transport of spin helical Dirac fermions being robust to disorder, our findings could also be important in the search for Majorana fermions in quantum wires \cite{Juan2014,Cook2011,Cook2012,Ilan2014}.

\vspace{-5mm}
\subsection*{Acknowledgments} 
\vspace{-5mm}
J.D. acknowledges the support of the German Research Foundation DFG through the SPP 1666 "Topological Insulators" program.




\end{document}


\title{Pseudo-ballistic transport \\in 3D topological insulator quantum wires :\\ Supplementary Information}

\author{Joseph Dufouleur$^1$, Louis Veyrat$^1$, Emmanouil Xypakis$^2$, Jens~H. Bardarson$^2$, Christian Nowka$^1$, Silke Hampel$^1$, Barbara Eichler$^1$, Oliver~G. Schmidt$^1$, Bernd B\"uchner$^{1,3}$, and Romain Giraud$^{1,4}$}
\affiliation{
\vspace{5mm}
$^{1}$Leibniz Institute for Solid State and Materials Research, IFW Dresden, D-01069 Dresden, Germany\\
$^{2}$Max-Planck-Institut f\"ur Physik Komplexer Systeme, N\"othnitzer Stra{\ss}e 38, D-01187 Dresden, Germany\\
$^{3}$Department of Physics, TU Dresden, D-01062 Dresden, Germany\\
$^{4}$CNRS - Laboratoire de Photonique et de Nanostructures, Route de Nozay, F-91460 Marcoussis, France\\
}

\date{\today}


\maketitle

\section{Experiments}

\subsection*{Length dependence of non-universal conductance fluctuations}

\begin{figure*}[!t!h]
\includegraphics[width=1\columnwidth]{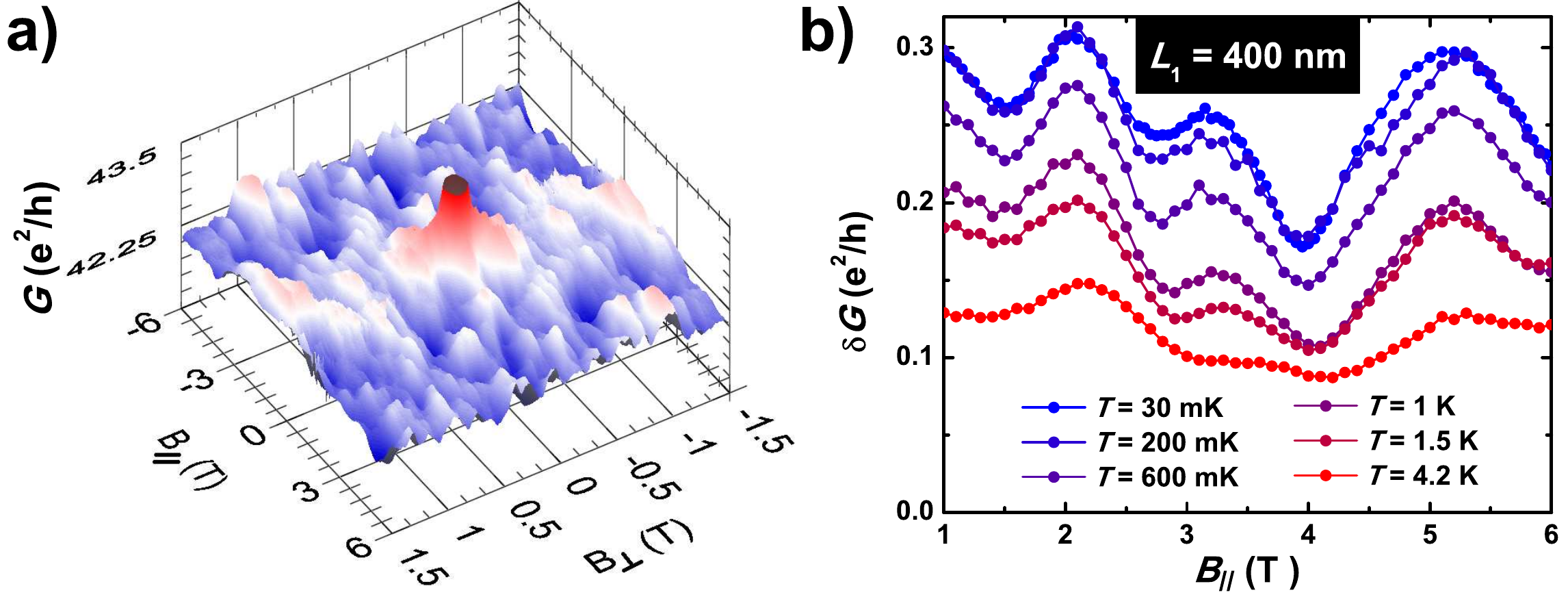}
\centering
\caption{
\textbf{a}, Mapping of the magneto-conductance of a quantum wire of length $L_1=400$~nm, with $L_1/l_\text{tr} \gtrsim 1$, measured at $T=30$~mK. 
\textbf{b}, Longitudinal-field dependence of the standard deviation of conductance fluctuations $\delta G$, measured at different temperatures. The modulation of the variance becomes temperature independent below about 1~K.
}
\label{fig1}
\end{figure*}

\begin{figure*}[!t!h]
\includegraphics[width=\columnwidth]{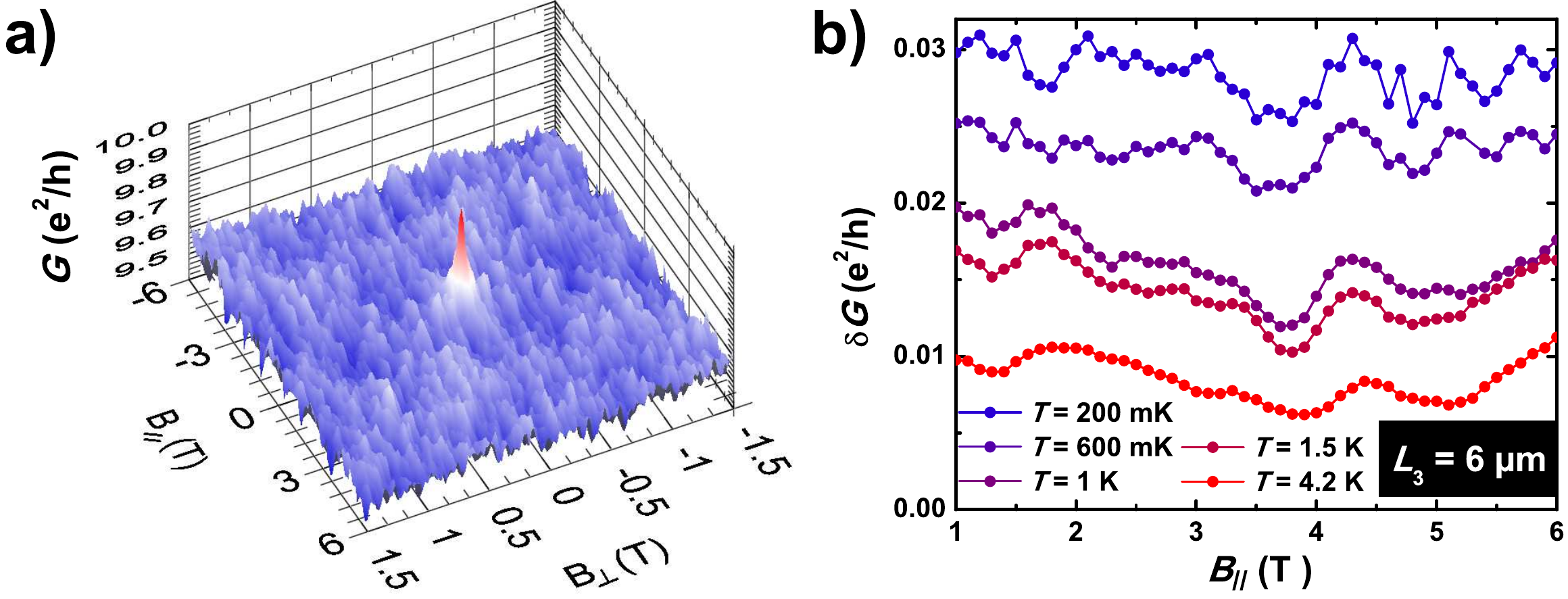}
\centering
\caption{
\textbf{a}, Mapping of the magneto-conductance of a quantum wire of length $L_3=6~\mu$m, with $L_3/l_\text{tr} \approx 20$, measured at $T=200$~mK.  
\textbf{b}, Longitudinal-field dependence of $\delta G$, measured at different temperatures. The modulation of the variance is strongly damped with respect to shorter wires, but it is still clearly visible and it shows a temperature dependence similar to the case of other lengths.  
}
\label{fig2}
\end{figure*}

All mesoscopic conductor studied have a length $L$ that is comparable to or much longer than the transport mean free path $l_\text{tr}$, so that the longitudinal motion of helical Dirac fermions is diffusive. As expected from theory for quantum coherent transport, the amplitude of conductance fluctuations is reduced in long wires, due to an averaging between uncorrelated coherent segments when $L > L_{\varphi}(T)$. The amplitude of non-universal conductance fluctuations is also reduced when the wire length is increased, but this already happens for $L \ge l_\text{tr}$. As further shown in Fig.~\ref{fig1} and Fig.~\ref{fig2}, a remarkable property of the modulation seen in $\delta G(B_{\parallel})$ is its relatively weak temperature dependence, even for the longest quantum wire. This is typical of pseudo-ballistic transport, and it clearly shows the importance of the large transverse energy quantization $\Delta$, which remains larger than the energy level broadening $\Gamma$. 
A direct consequence is that the phase coherence length does not directly control the temperature dependence of non-universal conductance fluctuations. Besides, this modulation has no more temperature dependence below $T^{\ast} \approx1$~K, for all lengths. Since the disorder broadening of energy levels $\Gamma$ should be independent of the length of a conductor when $L \ge l_\text{tr}$, the same crossover occurs when thermal broadening becomes smaller than $\Gamma$, which corresponds to our observations.

\section{Theory}

\subsection{Theoretical model}

To theoretically model our experiments we adapt a continuous Dirac fermion description of the surface state \cite{Bardarson2010}, and take the bulk to be an inert insulator. 
Explicitly, the surface Hamiltonian reads
\begin{equation}
\label{eq:Dirac}
H = v(\mathbf{p}+e\mathbf{A})\cdot\mathbf{\sigma} + V(\mathbf{r}),
\end{equation}
where $v$ is the Fermi velocity, $\mathbf{\sigma}=(\sigma_x,\sigma_y)$ are Pauli matrices, and the applied magnetic field $\mathbf{B} = \mathbf{\nabla}\times\mathbf{A}$. We take $\mathbf{r} = (x,y)$ with $x$ the direction along the length of the wire, and $y$ the periodic transverse direction. The spin of the Dirac fermion is constrained to lie in the tangent plane to the surface, and therefore rotates by $2\pi$ going once around the circumference of the wire. This leads to a Berry's phase of $\pi$ that is taken into account via the boundary condition
\begin{equation}
\label{eq:bc}
\psi(x,y+W) = \psi(x,y)e^{i\pi},
\end{equation}
with $W$ the wire circumference. Disorder is introduced through the time reversal invariant scalar potential $V$ with correlator
\begin{equation} 
\langle V(\mathbf{r})V(\mathbf{r}^\prime) \rangle = g(\hbar v)^2/(2\pi\xi^2)e^{-|\mathbf{r}-\mathbf{r}^\prime|^2/2\xi^2},
\end{equation} 
whereby $g$ is a dimensionless measure of the disorder strength and $\xi$ gives the characteristic length scale of potential variations. The Hamiltonian~\eqref{eq:Dirac}, together with metallic lead boundary conditions, defines a scattering problem that is solved via a transfer matrix technique \cite{Bardarson2007a}, giving the conductance through the Landauer-B\"uttiker equation. $\mu$ is the chemical potential in the wire.

\subsection{Transverse-field dependence of the quantized energy spectrum}

\begin{figure}[!t!h]
\includegraphics[width=\columnwidth]{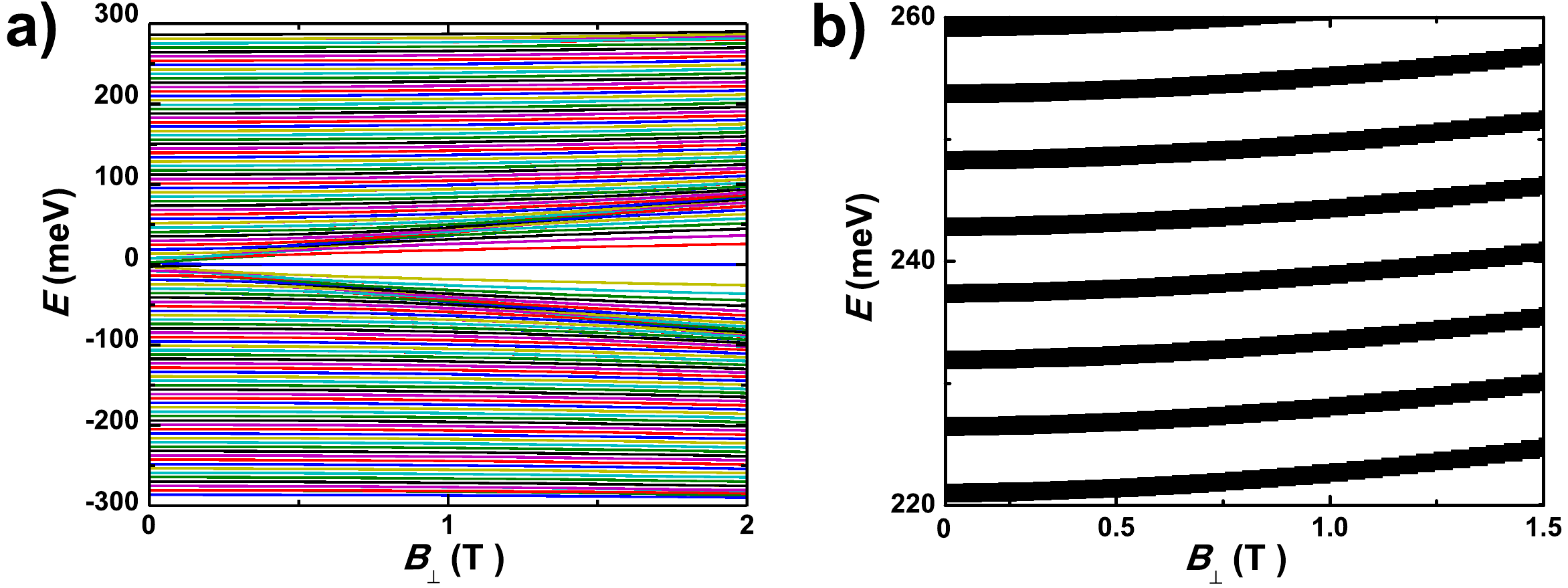}
\centering
\caption{
\textbf{a}, Transverse field dependence of the energy spectrum of a nanoribbon with a height $h=20$~nm and width $w=170$~nm.  
\textbf{b}, Zoom in the high-energy range, corresponding to the typical Fermi energy in highly-doped Bi$_2$Se$_3$ nanostructures.  
}
\label{fig3}
\end{figure}

As previously described in ref.~\cite{Juan2014}, we calculated the energy spectrum of a topological insulator quantum wire with a rectangular cross section (height $h=20$~nm, width $w=170$~nm).
Fig.~\ref{fig3} shows the dependence of the quantized energy spectrum with a transverse magnetic field $B_{\perp}$. 
Since a transverse magnetic fields breaks the initial symmetry of the Dirac Hamiltonian, it favors the mixing of transverse modes. 
Close to the Dirac point, this mixing can be so strong that edge states rapidly develop when $B_{\perp}$ is increased \cite{Juan2014}, as seen in Fig.~\ref{fig3}a). For a fixed value of $B_{\perp}$, the degree of mixing is reduced when the kinetic energy of quasi-particles becomes larger than the Zeeman energy. Therefore, $B_{\perp}$ has little influence on the spectrum of high-energy modes and, in the field and energy range studied (see Fig.~\ref{fig3}b), the slow increase in the transverse kinetic energy remains smaller than the level spacing $\Delta$.



\newpage 

\newpage 

\newpage 

